\documentclass[12pt,a4paper]{article}
\usepackage{authblk}  
\usepackage[utf8]{inputenc}
\usepackage[colorlinks=true, linkcolor=blue, citecolor=blue, urlcolor=blue]{hyperref}
\usepackage{graphicx}
\usepackage{amsmath}
\usepackage{amssymb}
\usepackage{cite}
\usepackage[margin=2.5cm, bottom=2cm]{geometry} 
\usepackage{parskip}

\title{Three-dimensional Meta-atoms for High Confinement of Mid-IR Radiation}

\author[1*]{Francesco Pisani}
\author[1]{Usama Iqbal}
\author[2]{Laure Tailpied}
\author[2]{Baptiste Fix}
\author[3]{Isabelle Sagnes}
\author[1]{Yanko Todorov}
\small{
\affil[1]{\small{Laboratoire de Physique et d'Étude des Matériaux, LPEM, UMR 8213, ESPCI Paris, Université PSL, CNRS, Sorbonne Université, F-75005 Paris, France}}
\affil[2]{DOTA, ONERA, Université Paris-Saclay, F-91123 Palaiseau, France}
\affil[3]{Centre de Nanosciences et de Nanotechnologies, CNRS – Université Paris-Sud-Paris-Saclay, 1 avenue Augustin Fresnel, 91120 Palaiseau, France}
\affil[*]{francesco.pisani@espci.fr}
}

\begin{document}
\maketitle
\begin{abstract}
The ability to confine photons into structures with highly sub-wavelength volumes is extremely interesting for many applications such as sensing, nonlinear optics, and strong light-matter interactions. However, their realization is increasingly difficult as the wavelength becomes shorter, due to fabrication challenges and increased metal losses. In this work, we present the first experimental characterization of three-dimensional circuit-like resonators operating in the mid-infrared. 
Through a combination of simulations, reflectivity measurements, and scanning near-field optical microscopy, we developed an analytical model capable of predicting the electromagnetic response of these structures based on their geometrical parameters. The design we studied offers a high degree of flexibility, enabling precise control over the resonant frequency of the various modes supported by the resonator, as well as independent control over radiative and non-radiative losses. Combined with the extreme field confinement demonstrated, these meta-atoms are highly promising for applications in detectors, emitters, nonlinear processes, and strong light-matter coupling.

\end{abstract}

\section{Introduction}
Metamaterials have emerged as a powerful tool for manipulating and confining electromagnetic waves, from the GHz domain to the visible part of the spectrum \cite{Chen2016,Capolino2009}. 
By tailoring the subwavelength structural units and their periodic arrangement, metamaterials achieved properties like negative refractive indexes \cite{Collin2014,Zhang2005}, tunable dielectric and magnetic permittivity \cite{Forouzmand2019,He2015,Pendry1999} or flat lenses \cite{Khorasaninejad2016}. While the collective response of the array defines the metamaterial's overall behavior, the individual structural units, known as meta-atoms resonator, can be freely engineered to exhibit the desired behavior.
Of particular interest are three-dimensional meta-atoms, where resonators are designed to confine electromagnetic waves in all three spatial dimensions. This geometry is typically achieved by combining a metamaterial concept in multilayered systems. The dielectric medium, hosting the electric field, is sandwiched between two metallic layers. The stack of metal-dielectric-metal is usually referred as metal-metal (MM) cavity \cite{nga2014antenna,bigioli2020long,pisani2023electronic}. Both the top and bottom layers can be patterned to achieve the desired electromagnetic response. Common design for these meta-atoms includes ridges, patches or inductor-capacitor (LC) resonator \cite{jeannin2019ultrastrong}.
Owing to their three-dimensional design, these meta-atoms confine light within highly sub-wavelength volumes and generate exceptionally high field intensities. LC resonators, in particular, enable quasi-static operation, where electric and magnetic fields oscillate in spatially distinct regions: in the inductor and in the capacitor respectively. 
The capacitive volume can be made extremely small without changing the resonant frequency and, provided that a good optical absorption is maintained, this miniaturization results in the confinement of the electric field in extraordinarily sub-wavelength volumes, leading to significantly enhanced field intensities.
This geometry is especially promising for various applications, as the dielectric layer can incorporate active regions, such as semiconductor heterostructures. Potential applications include enhanced light emission \cite{walther2010microcavity}, increased antenna effect for high responsivity, low dark current photodetectors  \cite{Jeannin2020a}, non-linear frequency conversion or harmonic generation \cite{Lee2014,yu2019third}, and strong field localization for surface-enhanced spectroscopy \cite{Banbury2019} and strong light-matter coupling \cite{ciuti2005quantum}.
For all these applications it is essential to control not only the resonant frequency of the meta-atom but also its radiative and non-radiative losses and the light-matter coupling of the structure as a whole. 

Three-dimensional meta-atoms are relatively straightforward to fabricate in the low-frequency part of the spectrum, as their micrometer-sized features can be easily fabricated using modern nanofabrication techniques. For instance, the terahertz frequency range ($\lambda=300$ $\mu$m$-30$ $\mu$m), has seen a flourishing of various three-dimensional designs \cite{walther2010microcavity, Todorov2015, Mottaghizadeh2017,Jeannin2020b,Jeannin2020c, Dietze2013, Paulillo2014a, Paulillo2016, Aupiais2023,Messelot2020}. On the contrary, in the mid-infrared (MIR) range ($\lambda=30$ $\mu$m$-3$ $\mu$m), where feature sizes are an order of magnitude smaller, mainly patch antennas \cite{palaferri2018room,Rodriguez2022a} and their derivations \cite{Lee2014} have been implemented, but they operate in a propagating mode. Quasi-static meta-atoms require much smaller dimensions than resonators based on propagation. So far, such structures have been reported in the mid-infrared range as exclusively planar structures \cite{benz2015control}. \\

In this work, we explore the concept of three-dimensional LC circuit resonators in the MIR and, to the best of our knowledge, present the first experimental characterization of such structures.
We provide a systematic experimental study of the resonator optical properties by varying a large number of geometrical parameters of the meta-atom. These spectroscopic measurements are complemented by numerical simulations and Scanning Near-Field Optical Microscopy (SNOM) measurements. Notably, we confirm the quasi-static behavior of the fundamental resonance in our structures. \\
By combining experimental results with simulations, we developed a general analytical model that describes the interaction of meta-atoms with free space, accurately predicting the resonant frequency and losses of the LC mode as a function of various geometrical parameters. Additionally, we demonstrate that this geometry offers remarkable flexibility in independently tuning the first, second, and third-order resonant modes.\\
The large degrees of freedom to engineer radiative and non-radiative losses of the structures, the extremely sub-wavelength volume of the resonance, and the high tunability of higher order modes make this platform highly relevant for applications in detectors, emitters, nonlinear effects, and strong light-matter coupling. 

\section{Results}
\begin{figure}[ht]
    \centering
    \includegraphics[width=.85\textwidth]{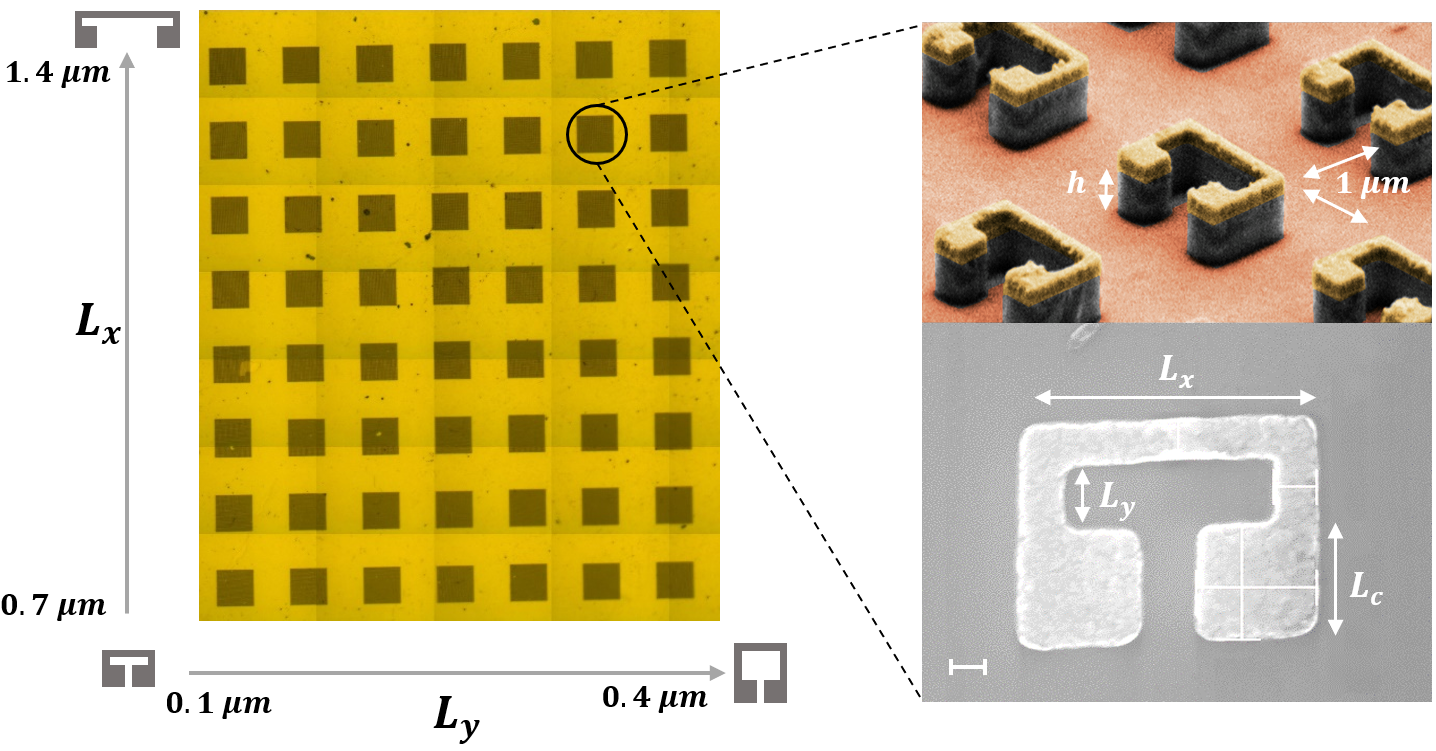}
    \caption{Microscope picture of the fabricated sample. It consists of an $8\times7$ matrix; each element is a $100\times100$ $\mu$m$^2$ array of gold meta-atom resonators with different geometrical aspect (see inset). Along the rows we changed the length of the horizontal inductive part ($L_x$) while on the columns we changed the vertical section ($L_y$). The size of the capacitor part ($L_C$) is kept constant. The resonators are evaporated on top of a wafer-bounded gallium arsenide layer of thickness $h=150$ nm and $300$ nm thickness. The inset shows two electron microscope images of one of the arrays, illustrating the geometrical parameters. The scale bar in the bottom-left corner represents 100 nm.}
    \label{Fig1}
\end{figure}

\textbf{Sample description.} The devices we fabricated (Fig.\ref{Fig1}) consist of an array of subwavelength resonators designed as three-dimensional meta-atoms. The inset in Fig. \ref{Fig1} displays a Scanning Electron Microscope (SEM) image, presented in false colors. The colored sections (bottom and top parts of the resonator) represent gold, while the gray area corresponds to gallium arsenide (GaAs). Detailed fabrication procedures are provided in the \textbf{Methods} section.
The resonator is composed of two squared patches of size $L_C^2$ and a wire connecting them of length $L_x+2L_y$. To thoroughly characterize the resonance behavior of this structure, we fabricated multiple arrays with variations in the wire length by independently modifying both $L_x$ and $L_y$. An additional set of arrays where we varied the patches size, $L_C$, was fabricated for completeness (see Supporting Information). The thickness of the GaAs was also varied: $h=150$ nm and $300$ nm, to assess its influence on the resonance.\\ 
The reflectivity of each array was measured with a micro-FTIR (see methods for additional information) for both $x$- and $y$-polarized light; $x$ and $y$ are defined as the axes parallel to $L_x$ and $L_y$ respectively.

\begin{figure}[ht]
    \centering
    \includegraphics[width=1\textwidth]{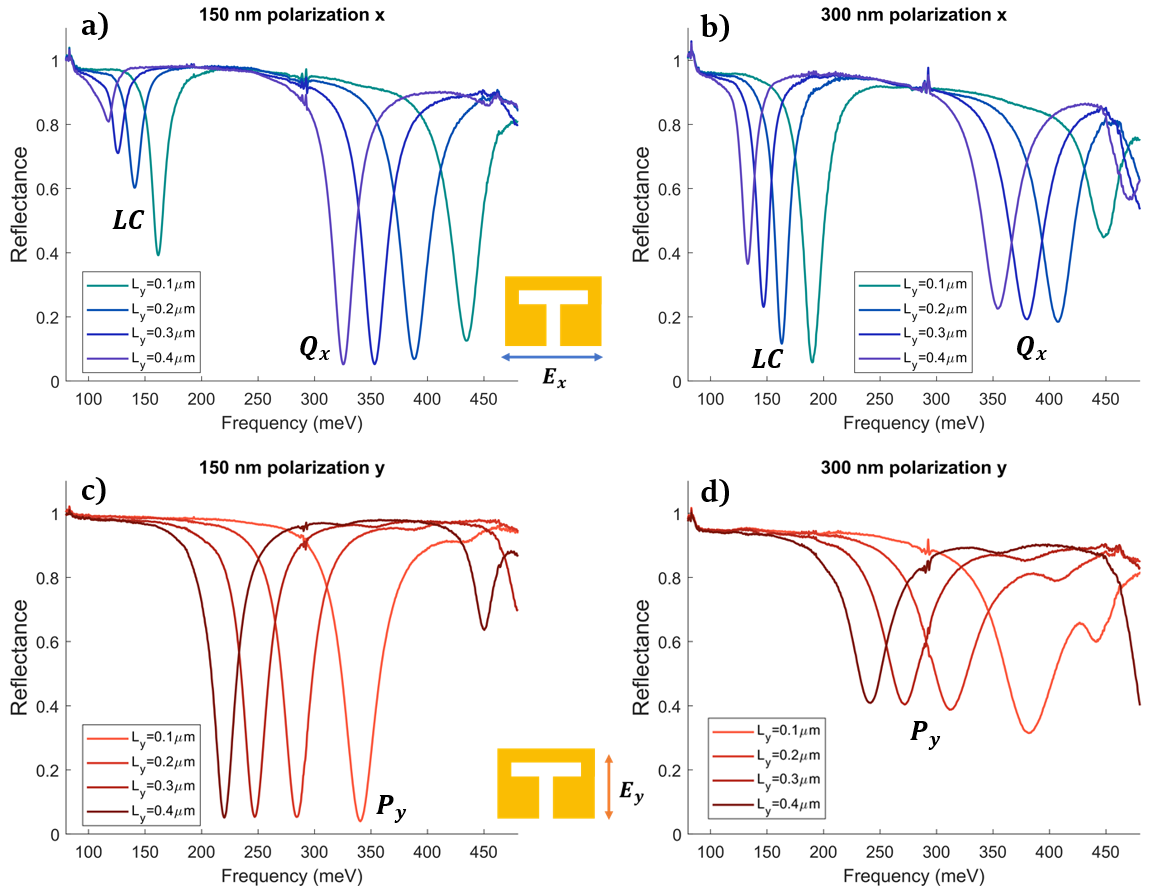}
    \caption{Reflectivity spectra, normalized to a gold background, are shown for a selected set of parameters: $L_x=0.8$ $\mu$m and  $L_y=0.1,0.2,0.3,0.4$ $\mu$m. The spectra correspond to samples with GaAs layers of 150 nm thickness (a, c) and 300 nm thickness (b, d). The spectra in blue (a, b) are taken with x-polarized incident light, while the one in red (c, d) with y-polarized light. In the $x$-polarization, the lower-energy peak corresponds to the LC mode, while a higher-energy peak is observed, associated with a quadrupolar field distribution. For the $y$-polarization, the observed peak corresponds to a mode resembling the behavior of patch cavities. A detailed view of the field distributions for each mode is provided in Fig. \ref{Fig3}.}
    \label{Fig2}
\end{figure}

 In the x-polarization, Fig. \ref{Fig2} a, b, we observe two absorption peaks in the spectral band of the FTIR. The low frequency one, which spans between $100-200$ meV, is the fundamental mode supported by this structure and is akin to "LC" (Inductor-Capacitor) resonators where the electric field is stored in the capacitors, the two squared patches, and the magnetic field in the wire (in the supporting information we report a simulation that highlights this behavior).\\
 At a frequency roughly 3 times higher, we observed a second peak named “$Q_x$” (Quadrupolar mode). In the $y$-polarization we observe a different peak of absorption, “$P_y$” (Patch mode), roughly at the twice the frequency of the fundamental mode (Fig.\ref{Fig2} c, d). The choice of these names comes from the field distribution and will be clarified later (see \textbf{Modes field distribution}). Higher-order modes are also observed in both $x$- and $y$-polarizations; however, the discussion here is limited to these three resonances for the sake of conciseness. The data presented correspond to a single row of the matrix design, where $L_x=0.8$ $\mu$m for different values of the length of $L_y$. This selection highlights the spectral behavior of the structures and the trends as a function of geometry (see Supporting Information for the full characterization). In particular, as the length of the wire  increases, the resonant frequency of the modes decreases.

The electromagnetic response of the structures, encompassing all combinations of $L_x$, $L_y$ and thicknesses $h$, was simulated with a commercial finite element electromagnetic solver, COMSOL. The simulated reflectivity closely matches the experimental data, demonstrating excellent agreement in terms of resonance frequency, linewidth, and reflectivity contrast. The complete dataset is provided in the Supporting Information.

\begin{figure}[ht]
    \centering
    \includegraphics[width=1\textwidth]{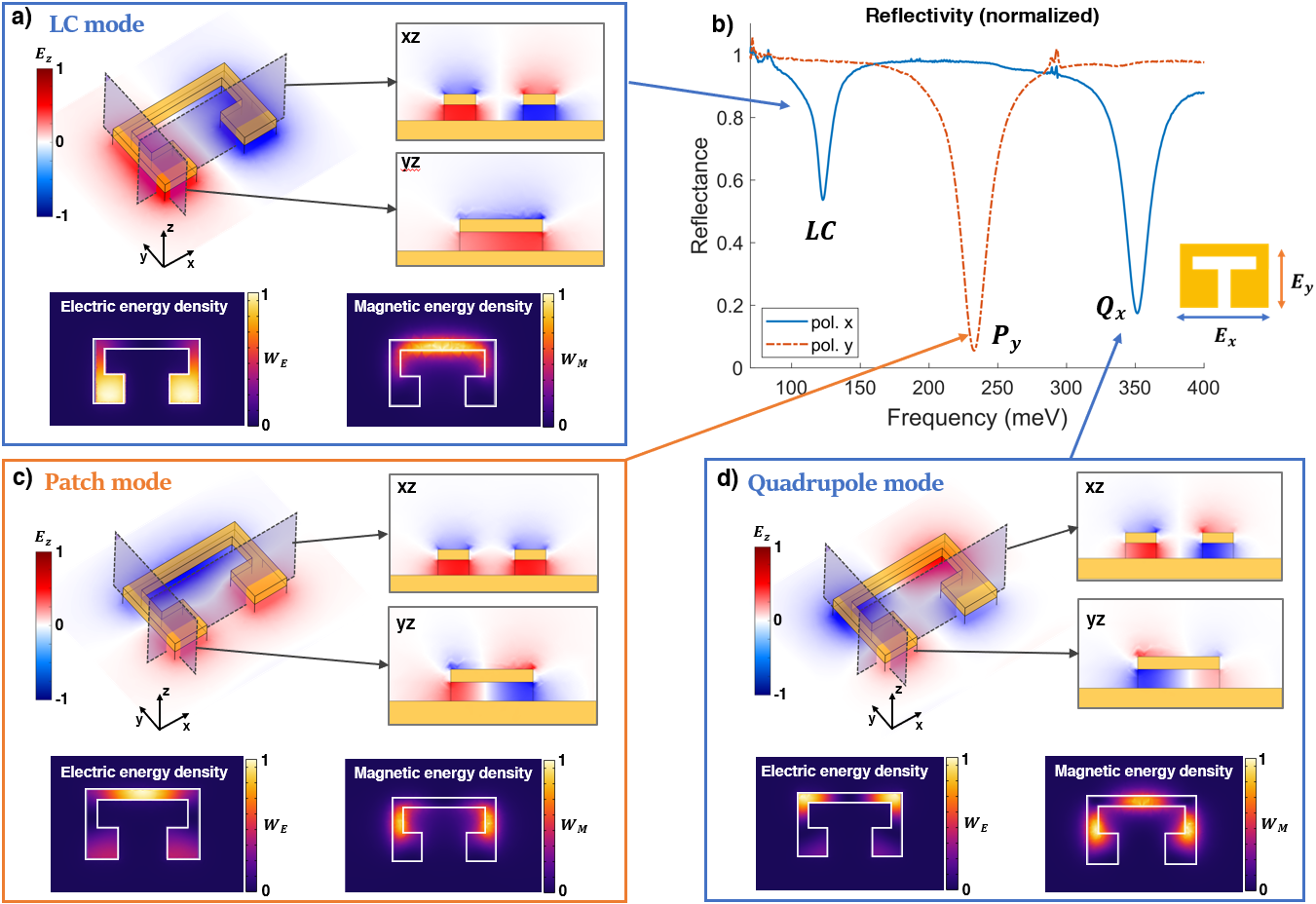}
    \caption{Electric field ($z$-component) distribution of the various resonant modes of the meta-atom. The first mode, observed for $x$-polarized light, corresponds to the LC mode (a). Here, the electric field is predominantly concentrated beneath the capacitors, while the magnetic field is localized around the inductive elements. The second-order mode, observed for $y$-polarized light, resembles a patch cavity mode, with the electric field distributed along the sides of the structure (c). A third-order mode, observed in the $x$-polarization, exhibits a quadrupolar distribution of both the electric and magnetic fields (d). The reflectivity dips corresponding to the simulated mode are reported in b). The fundamental LC mode and the quadrupolar mode ($Q_x$) are excited in the $x$-polarization (blue line), while the patch mode is excited in the $y$-polarization (orange line).}
    \label{Fig3}
\end{figure}

\textbf{Modes field distribution}. The numerical simulations also give access to the field distribution as shown in Fig.\ref{Fig3}. The first resonance, the “LC” mode, is depicted in Fig.\ref{Fig3}, a), where the z-component of the electric field is plotted in the $xy$, $xz$ and $yz$ planes (refer to Fig. \ref{Fig3} for the coordinate system definition). The observed electric field distribution resembles that reported for structures designed for the THz domain \cite{Jeannin2020c}.
Specifically, the electric energy density is concentrated beneath the two squared patches, the capacitive sections, while the magnetic energy density is localized around the wire, the inductive element. This separation of fields is characteristic of LC modes, where the resonator size is much smaller than the wavelength, allowing the quasi-static approximation to describe them \cite{Jeannin2020c}. This assumption will be further supported by the analytical modeling of the resonant frequency we will present later.
The structure also supports resonant modes for $y$-polarized light, primarily influenced by the size along the $y$-axis. By looking at the $yz$-plane cross-section shown in Fig.\ref{Fig3} c, one can see the field intensity showing a sinusoidal distribution. This behavior is typical for patch cavities \cite{palaferri2018room,Rodriguez2022a,todorov2010optical}.

In Fig.\ref{Fig3}, b), we report the reflectivity spectra corresponding to each field distribution; particularly interesting is the second order in the $x$-polarization (Fig.\ref{Fig3}, d). The electric field for this mode exhibits four maxima located at the edges of the resonator, consistent with the field distribution of a quadrupolar mode. Typically, such modes are dark in patch cavities due to their lack of a net dipole moment \cite{todorov2010optical}. Under these conditions, they are observable only in simulations or at large angles of incidence. 
However, in our resonator, the quadrupolar mode acquires an effective net dipole moment due to the asymmetry of the top metallic layer. Specifically, Fig.\ref{Fig3}, d) shows that the $E_z$
field at the edges of the square loop is significantly stronger than the corresponding field components in the two capacitive pads. This asymmetry generates a net dipole moment, which is responsible for the observed absorption peak in the reflectivity spectrum, as we will discuss in more detail.
 
\begin{figure}[ht]
    \centering
    \includegraphics[width=1\textwidth]{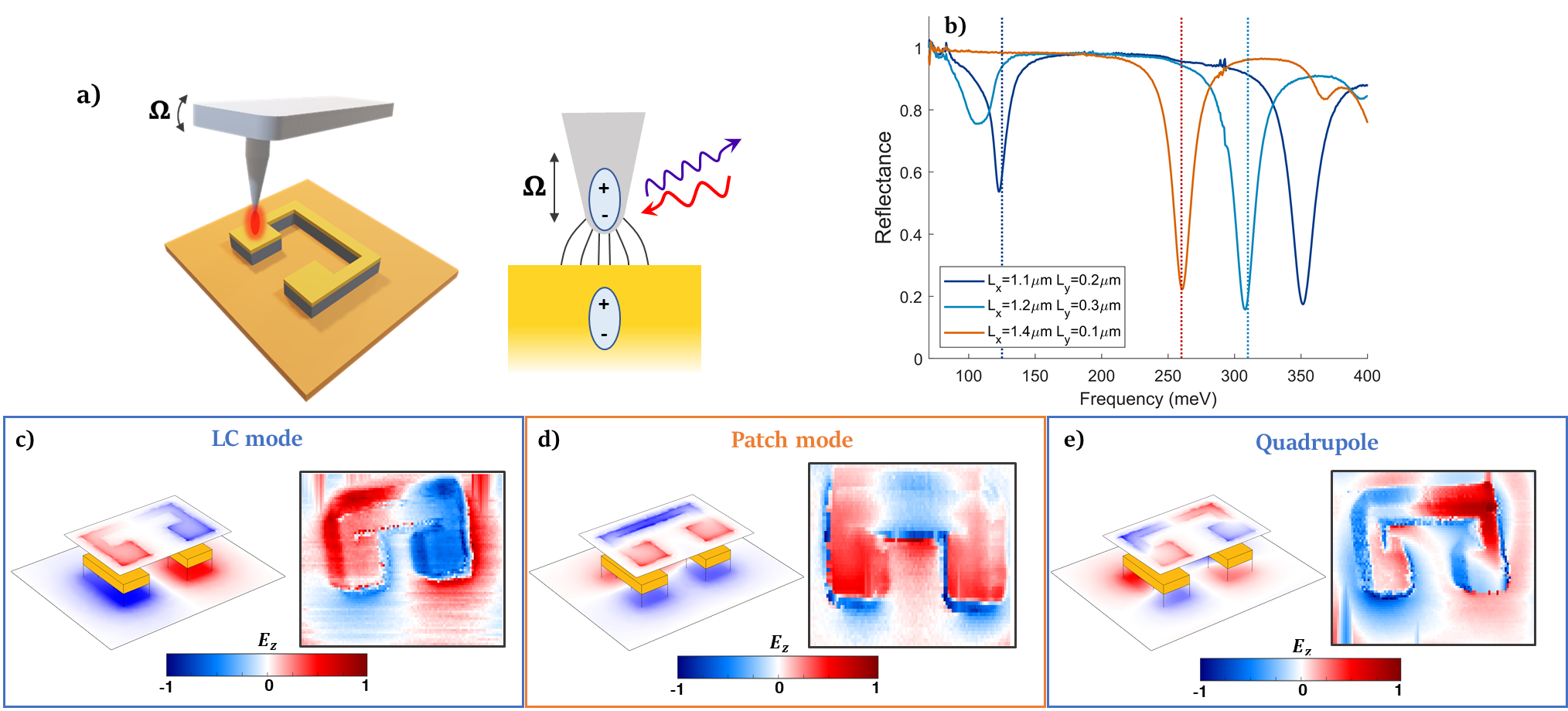}
    \caption{Scheme of a scanning near optical microscopy setup a). The reflectance spectra of the three arrays selected for measurements are shown in (b), with dotted lines indicating the laser frequencies used in the experiments. Panels (c-e) display SNOM measurements for the three previously analyzed modes. The colormap represents the phase of the $z$-component of the electric field: LC mode (c), Patch mode (d), and Quadrupolar mode (e). The measurements closely match the simulations. Notably, the abrupt phase change occurs at the resonator edges, where the electric field changes sign, as also depicted in Fig. \ref{Fig3}.}
    \label{Fig4}
\end{figure}

The field distribution of the meta-atom was experimentally measured using a Scanning Near-Field Optical Microscope (SNOM), Fig.\ref{Fig4}, a) \cite{thomas2022imaging,neuman2015mapping}. See Methods for details regarding the measurement setup. The results are presented in Fig.\ref{Fig4}, c-e), where the phase of the $z$-component of the electric field, measured with the SNOM (insets), is compared to simulations for the three resonant modes. Each panel shows the field distribution corresponding to the absorption peaks reported in Fig.\ref{Fig4}, b). Notably, because the SNOM maps the field just above the structure, the field inversion observed in the $xz$ and $yz$ planes of Fig.\ref{Fig3} is also detected. Specifically, when the SNOM tip moves from the top of the resonator to the bottom metallic plate, the field undergoes a sign change. By simulating the field both on the top and bottom of the resonator, we successfully reproduced the measurements performed with the SNOM, achieving excellent agreement between experiment and simulation.

\section{Data analysis: resonant frequency and losses}
The primary objective of this work is to develop a model that qualitatively and quantitatively describes the electromagnetic response of the three-dimensional LC resonators. This model aims to provide a robust framework for predicting the mode-specific resonant frequencies, radiative and non-radiative losses. These three quantities are key element to optimize the resonator for specific applications, e.g. maximizing the absorption for photo-detector schemes.\\
To achieve this, we fitted the absorption peaks observed in both experimental data and simulations using a Lorentzian function. This fitting process enabled the extraction of key parameters: the peak frequency, contrast ($C$), and linewidth ($\Gamma$). The contrast and linewidth, in turn, were used to calculate the radiative and non-radiative losses, based on the formulas \cite{Jeannin2020b,Rodriguez2022a}:

\begin{equation}
\gamma_\text{nr} = \frac{\Gamma}{2} \left( 1 + \sqrt{1 - C} \right), \quad
\gamma_\text{rad} = \frac{\Gamma}{2} \left( 1 - \sqrt{1 - C} \right).
\label{losses}
\end{equation}
It is worth noting that the two loss terms appear symmetrically in the equations, making the choice of the +/- sign arbitrary. Later on, we will explain our choice for the attribution of radiative and non-radiative loss. 

\begin{figure}[ht]
    \centering
    \includegraphics[width=1\textwidth]{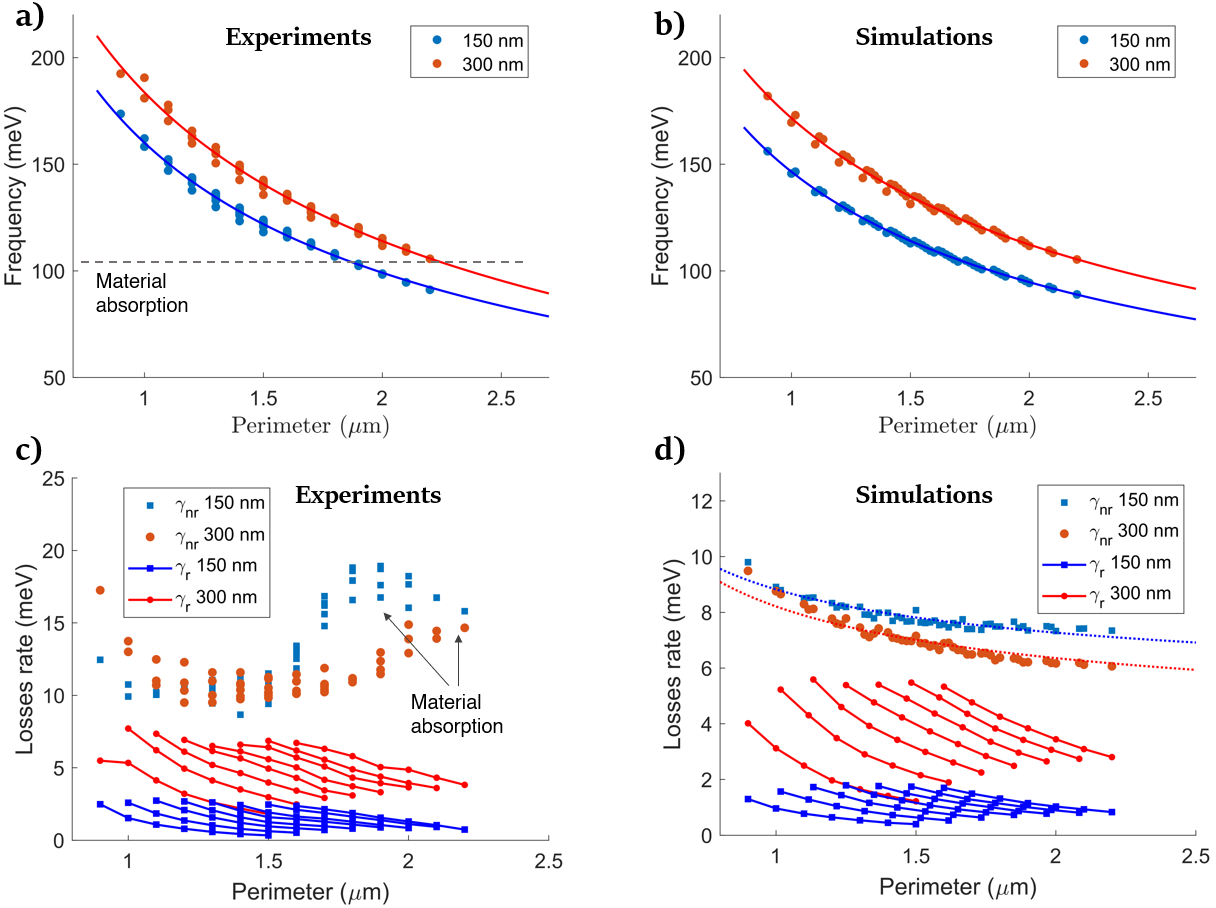}
    \caption{Resonant frequency of the LC mode as a function of the perimeter. Panels (a) and (d) show the resonant frequency of the LC mode for both simulations and experimental data, which align well with the theoretical prediction (continuous lines). Panels (b) and (e) display the radiative ($\gamma_r$) and non-radiative ($\gamma_{nr}$) losses as functions of the perimeter. In panel (e), the dotted lines represent the theoretical prediction for $\gamma_{nr}$. The data in panel (b) are influenced by additional material absorption in the GaAs near 105 meV, leading to an increase in $\gamma_{nr}$ for devices resonating around that frequency (highlighted by the dashed line in panel a).}
    \label{Fig5}
\end{figure}

\textbf{LC mode}. Guided by a quasi-static vision of the fundamental resonance, we model the structure as an effective LC circuit. The $\Pi$-shaped link connecting the two metallic pads is treated as a wire with a self-inductance given by:$ L = \left( \mu_0/2\pi\right) P \left( \ln\left( \frac{2P}{a} \right) - 1 \right)$, where $a$ is a geometrical parameter related to the wire thickness \cite{rosa1908self}, and $P=2L_x+L_y$ represents the perimeter of the inductive loop. The equivalent capacitance of the structure, $C_{eq} =C_{pp}/2+C_f$ accounts for two parallel capacitors in series, $C_{pp} = \varepsilon\varepsilon_0w^2/2h$, and an additional contribution from the fringing field, $C_f$,  which considers field leakage along the inductive elements and cross-capacitance between the two top square pads. The lumped element quantities $L$ and $C_{eq}$ provide an expression for the resonant frequency:
\begin{equation}
    \omega_{LC} = \frac{\sqrt{2\pi}}{\sqrt{C_{eq}\mu_0P\left[\ln\left( \frac{2P}{a} \right) - 1 \right]}}.
    \label{omegaLC}
\end{equation}
We used this formula to fit the data from experiments and simulations and extract the values of $C_{eq}$ and $a$. The resonant frequency of all devices was reported in figure \ref{Fig5}, a, b) for experiment and simulations. The continuous lines represent the fit of Eq.\eqref{omegaLC}. 
As a result, we obtain the parameter $a$ to be $70$ nm for the structure with $h=150$ nm and $65$ nm for $h=300$ nm. These values are obtained with a formula for a cylindrical wire (circular section) and need to be corrected by a factor $\sqrt{\pi}$ to account for the rectangular cross-section of our wire \cite{rosa1908self}. With this correction we find $a^*=124$ and $115$ nm, which are very similar for both structures and closely match the width of the metal wire connecting the two capacitors, as observed in SEM images (110 nm, Fig.\ref{Fig1}).\\
For the equivalent capacitance we obtain the values $C_{eq}= 35.3$ aF and $26.1$ aF for respectively $h=150$ nm and $h=300$ nm. Substracting the contribution of the two parallel plate capacitors in series, $C_{pp}/2= 28.2$ aF ($h=150$ nm) and $14.1$ aF ($h=300$ nm), we isolate the contribution from the fringing fields: $C_f = 7.1$ aF and $C_f = 12$ aF. For thicker structures, we expect the field to be partially delocalized along the $L_y$ (Fig.\ref{Fig3}), leading to a larger capacitance $C_f$ \cite{Jeannin2020c}.
The solid lines in Fig.\ref{Fig5}, panel (a), represent plots of Eq.\eqref{omegaLC} using the fitting parameters described above. It is remarkable that all data points align closely with the theoretical curves, despite significant variations in $L_x$ and $L_y$. 

The losses, extracted from the reflectivity data, were also reported in Fig.\ref{Fig5}, c, d). We identify the parameter $\gamma_{nr}$ (Eq.\eqref{losses}, left) as the ohmic loss of the resonators. Indeed, we observe that this parameter is relatively independent from the thickness $h$ and depends solely on the perimeter $P$ (not connected markers in Fig.\ref{Fig5}, b, e). This is consistent with the fact that ohmic losses mainly arise in the upper connecting wire, which is identical for both thicknesses. To confirm this interpretation, the resonator is modeled as an RLC circuit where the loss rate is provided by:
\begin{equation}
    \gamma_{nr}=R/L=\left(2\pi\rho/\mu_0\sigma\right)\left[\left( \frac{2P}{a} \right) - 1 \right]^{-1}
\end{equation}
where $\rho$ is the resistivity of the metal and $\sigma$ is the cross section for the current flow. The dotted lines in Figure \ref{Fig5}, d) correspond to the plot of $\gamma_{nr} (P)$ where we fitted the metal resistivity $\rho$ to match the data from the simulations. In order to be consistent with the previous fit, we used the previously determined value of $a$.  By fixing the wire cross-section $\sigma=1.2\times10^{-2}$ $\mu$m$^2$ (from SEM characterization) we obtained a gold resistivity $\rho=1.6\times10^{-8}$ $\Omega\cdot$m, close to the values reported to bulk gold: $2.2\times10^{-8}$ $\Omega\cdot$m, \cite{Matula1979}. The slightly lower value can be attributed to the very thin nature (110 nm) of the metal wire of the resonator. Another possible cause is overestimating the length of the resistive path, where the current flows. This hypothesis is supported by the simulation we performed: the current density is almost homogeneous along the wire length, with the exception of the section where the electric field leaks into the inductive wire, close to the capacitive sections, effectively decreasing the length of the current path $P$ where energy is dissipated. 

The experimental data revealed an additional source of non-radiative losses, particularly evident in the 150 nm thick sample. This absorption feature is consistently observed across all samples around 105 meV. 
We attribute this effect to the absorption caused by other chemical species incorporated into the GaAs layer during the MOCVD growth process, such as carbon impurities. These impurities, often originating from incomplete pyrolysis of the metal-organic precursors during growth, can act as non-radiative recombination centers, impacting material quality \cite{chen2020quantitative}.\\ 

\begin{figure}[ht]
    \centering
    \includegraphics[width=1\textwidth]{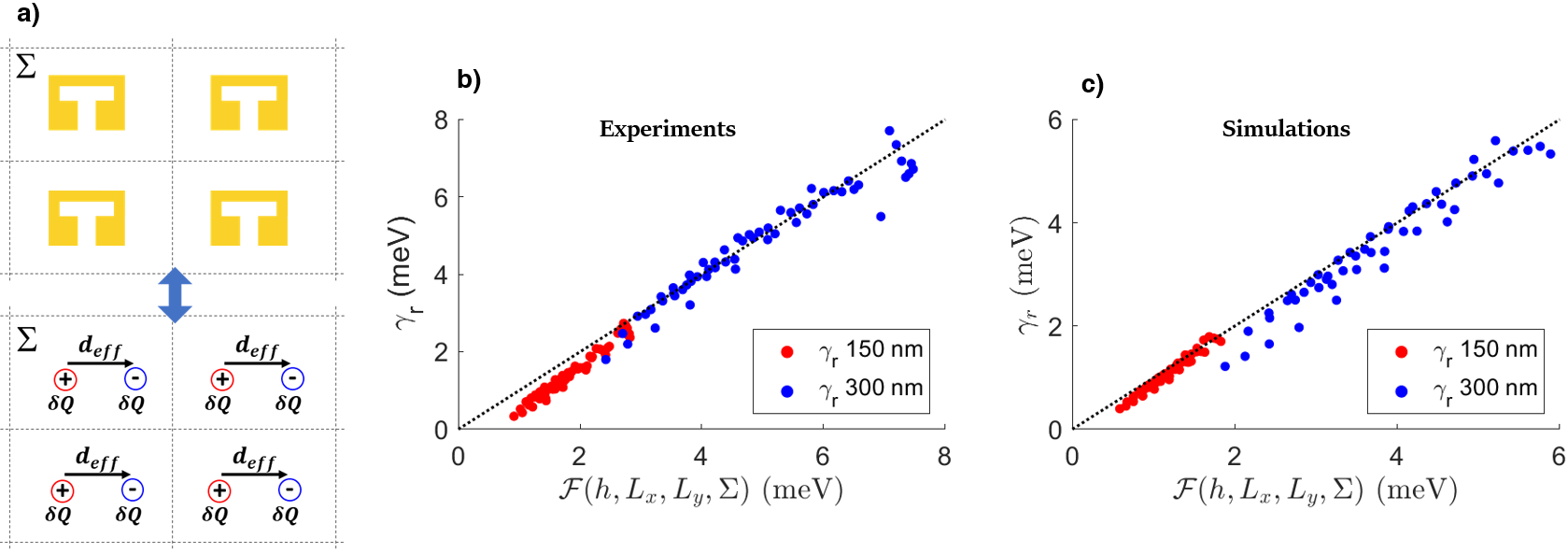}
    \caption{a) The complex geometrical shape of the meta-atom resonator is captured by an effective dipole $d_{eff}$, related to the width of the resonator and the induced charge $\delta Q$. b) Radiative losses from the experimental data c) and simulations d), are plotted as a function of $F(L_x,L_y)$ as described in the text. The dotted lines are guide for the eyes representing the $y=x$ line.}
    \label{Fig6}
\end{figure}

\textbf{Radiation loss of the LC mode.} The radiative losses exhibit a more complex dependence on the geometrical parameters of the metamaterial (Fig.\ref{Fig5}, b, e). A clear trend is the increase of $\gamma_r$ with the sample thickness $h$, which is a typical behavior for aperture antennas, such as patch antennas \cite{Orfanidis2024}. However, unlike in patch antennas, where $\gamma_r$ depends only on the thickness of the sample \cite{Rodriguez2022a, feuillet2013strong} and the array unit cell $\Sigma$, for this mode the radiation loss displays a qualitatively different behavior. In particular, $\gamma_r$ has a complex dependence on the parameters $L_x$ and $L_y$.\\
To develop a physical interpretation of these observations, we propose a general model for radiation losses applicable to any metamaterial geometry. This model provides insights into both quasi-static and propagating modes in resonators, offering a comprehensive framework for analyzing the electromagnetic response of these structures.
We assume that each element of the array can be modeled as an effective horizontal dipole $p_x = \delta Qd_{eff}$, where $d_{eff}$ is the length of the dipole and $\delta Q$ is the net charge induced on the left and right sections of the resonator. By searching for a solution of Maxwell equations for a periodic arrangement of dipoles and employing the Poynting theorem we were able to express the reflectivity coefficient $R=|E_r|^2/|E_{in}|^2$ . This relates to the radiation and non-radiation losses: $R=(\gamma_{rad}-\gamma_{nr})^2/(\gamma_{rad}+\gamma_{nr})^2$, and hence we deduce the following formula (derived in the supporting information):
\begin{align}
    \gamma_{\text{rad}} &=  \frac{d_{eff}^2}{\Sigma}\frac{ 2\pi c}{P} 
    \left[ \ln\left(\frac{2P}{a}\right) - 1 \right]^{-1}=F(h, L_x, L_y, \Sigma).
    \label{gammaradF}
\end{align}
Here $d_{eff}$ is an effective dipole associated to the geometrical shape of the resonator (Fig.\ref{Fig6}, a), and it can be shown to be proportional to the substrate thickens and $L_x$: $\displaystyle d_{eff}=\frac{1}{\sqrt{C_0}l_0}h L_x$. $C_0=C_{eq}/C_{pp}$ was determined before to be $1.25$ for $h=150$ nm and $1.85$ for $h=300$ nm, while $l_0$ is an unique constant $l_0\simeq3$ $\mu$m consistent across all combinations of $L_x$, $L_y$ and thicknesses found by fitting $d_{eff}$. See the Supporting Information for more details.

The parameter $\gamma_{rad}$ is now expressed solely in terms of the geometrical parameters of each structure. In Fig.\ref{Fig6} (c, d), we plot the radiative losses from both simulations and experimental data as a function of the right-hand side of Eq.(\ref{gammaradF}), with all quantities converted from frequency units into meV. The alignment of the data along the $y=x$ line for the LC mode demonstrates the accuracy of our approach.\\
The model successfully captures the dependence of both radiation and non-radiation losses on the full range of parameters employed, including $L_x$, $L_y$, and the thickness $h$, allowing the prediction of resonator losses directly from its geometry. For instance, Eq.\eqref{gammaradF} reveals that, for a fixed perimeter length, lower losses are achieved for smaller $L_x$. This means that one could fix the resonant frequency, determined solely by the perimeter (Eq.\eqref{gammaradF}), and independently manipulate the radiative coupling to external radiation by adjusting the aspect ratio $L_x/L_y$. Such flexibility enables, for example, the achievement of the critical coupling condition, where radiative and non-radiative losses are equal. It is also important to highlight that the presented model is general and can be extended to other resonators, where their complex geometries can be simply recast into the effective dipole representation (Fig.\ref{Fig6}, a).

\begin{figure}[ht]
    \centering
    \includegraphics[width=1\textwidth]{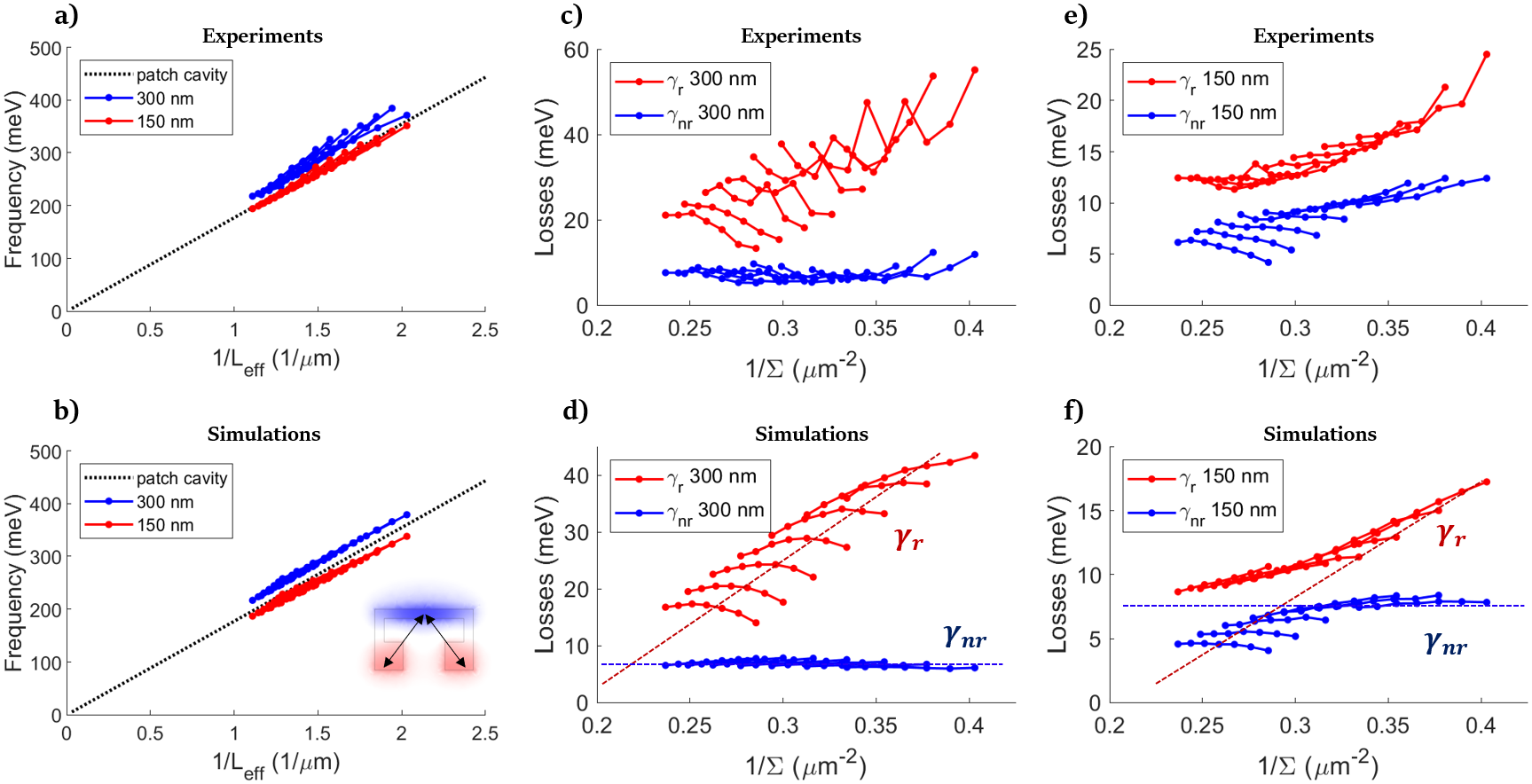}
    \caption{Resonant frequency of the patch-like mode as a function of the effective length $L_{eff}$ of the patch, a, b). Experiments (a) and simulations (b) show a linear behavior. The dotted lines represent the resonant frequency of a patch with a $n_{eff}=3.5$. Radiative and non-radiative losses of the structure with a thickness $h=300$ nm (c, d) and $150$ nm (e, f) as a function of the array unit cell area $\Sigma$. The non-radiative losses are almost constant (blue dashed lines) while the radiative ones are proportional to $h/\Sigma$. For the thinner structure we observe a crossing of the two, corresponding to the “critical coupling” condition. The dashed lines serve as a guide for the eyes, highlight the crossing between the two losses.}
    \label{Fig7}
\end{figure}

\textbf{Patch mode analysis}. By changing to $y$-polarized light, we observed another resonance akin to the propagation mode in patch antennas \cite{feuillet2012extremely}. Unlike the quasi-static mode discussed earlier, this mode exhibits qualitatively different behavior, with its resonant frequency following the patch cavity resonance formula \cite{todorov2010optical,feuillet2012extremely}: 
\begin{equation}
    \omega_{patch} = \frac{2 \pi c}{2 n_{eff} S_y}
    \label{omegapatch}
\end{equation}
where $S_y$ is the size of the patch and $n_{eff}$ is the effective refractive index of the material where the electric field is distributed. The latter is calculated in experiments by measuring the resonant frequency with reflectivity measurements. In standard GaAs patches, with thicknesses above $500$ nm, $n_{eff}$ is roughly 3, independently of $h$ \cite{Rodriguez2022a}. 
In our case, estimating $n_{eff}$ is more intricate due to the resonator's complex geometry and the propagation of the field through multiple media with different refractive indices (air, GaAs, and gold). Nevertheless, this complexity can be addressed by numerically calculating $n_{eff}$ using the following expression:
\begin{equation}
    n_{eff}^2 = \frac{\int_{V_{sim}} D_z^2 \, d^3x}{\varepsilon_0 W_e}:
    \label{neff}
\end{equation}
the integral in the simulation volume $V_{sim}$ of the squared $z$-component of the dielectric displacement field, $D_z^2$, normalized by the total electric energy, $W_e$.With this approach, the effective refractive index represents an averaged refractive index of the materials involved in the resonance, weighted by the electric field distribution. This includes contributions from air, where the fringing field resides, and the fraction of gold penetrated by the field (within the skin depth). In our case, due to the sample being particularly thin, the portion of field penetrating inside of the metal is significant and as such the effective refractive index calculated with eq.\eqref{neff} is higher then that of standard patches: $n_{eff}\simeq3.75$ for the 150 nm structure and $3.2$ for the 300 nm one.\\ 
Another important difference with the standard patch antenna geometry arises from the field distribution: by looking at figure \ref{Fig3}, c) and figure \ref{Fig4}, d) we can see that maximum of the electric field is localized in the capacitor pads, while the minimum at the center of the $L_x$ wire. To account for this, we replaced $S_y$ in eq.\eqref{omegapatch} with an effective length, $\displaystyle L_{eff}=\left[(L_y+L_C)^2+\frac{(L_x-L_C)^2}{4}\right]^{1/2}$, which reflects the minimal distance between the electric field's maximum and minimum (inset in Fig. \ref{Fig7}, b).  In Fig.\ref{Fig7}, a, b), we reported the resonant frequency as a function of $1/L_{eff}$ for both thicknesses. The dotted line represents the plot of eq.\eqref{omegapatch} with an effective refractive index of $3.5$, reported to highlight the linear behavior of the data.

The losses, figure \ref{Fig7} (c-f), also show a clearly different behavior from the LC mode: the non-radiative losses remain approximately constant as the geometrical parameters are varied, consistent with expectations for patch cavities \cite{Jeannin2020c}; in contrast, the radiative losses scale as the thickness over the array unit cell area: $h/\Sigma$. Particularly interesting is the plot we obtained in figure \ref{Fig7} (e, f). For the thinner sample, the radiative losses intersect the non-radiative losses, achieving the condition known as "critical coupling" \cite{Aupiais2023}, where the metamaterial's reflectance drops to zero. Since the two values are interchangeable in eq.\eqref{losses}, when fitting these sets of data, it seems that the non-radiative losses decrease at large areas, but this actually corresponds to the radiative crossing the non-radiative ones; the dotted lines in figure \ref{Fig7} (d, f) highlight this behavior.  

\begin{figure}[h!]
    \centering
    \includegraphics[width=1\textwidth]{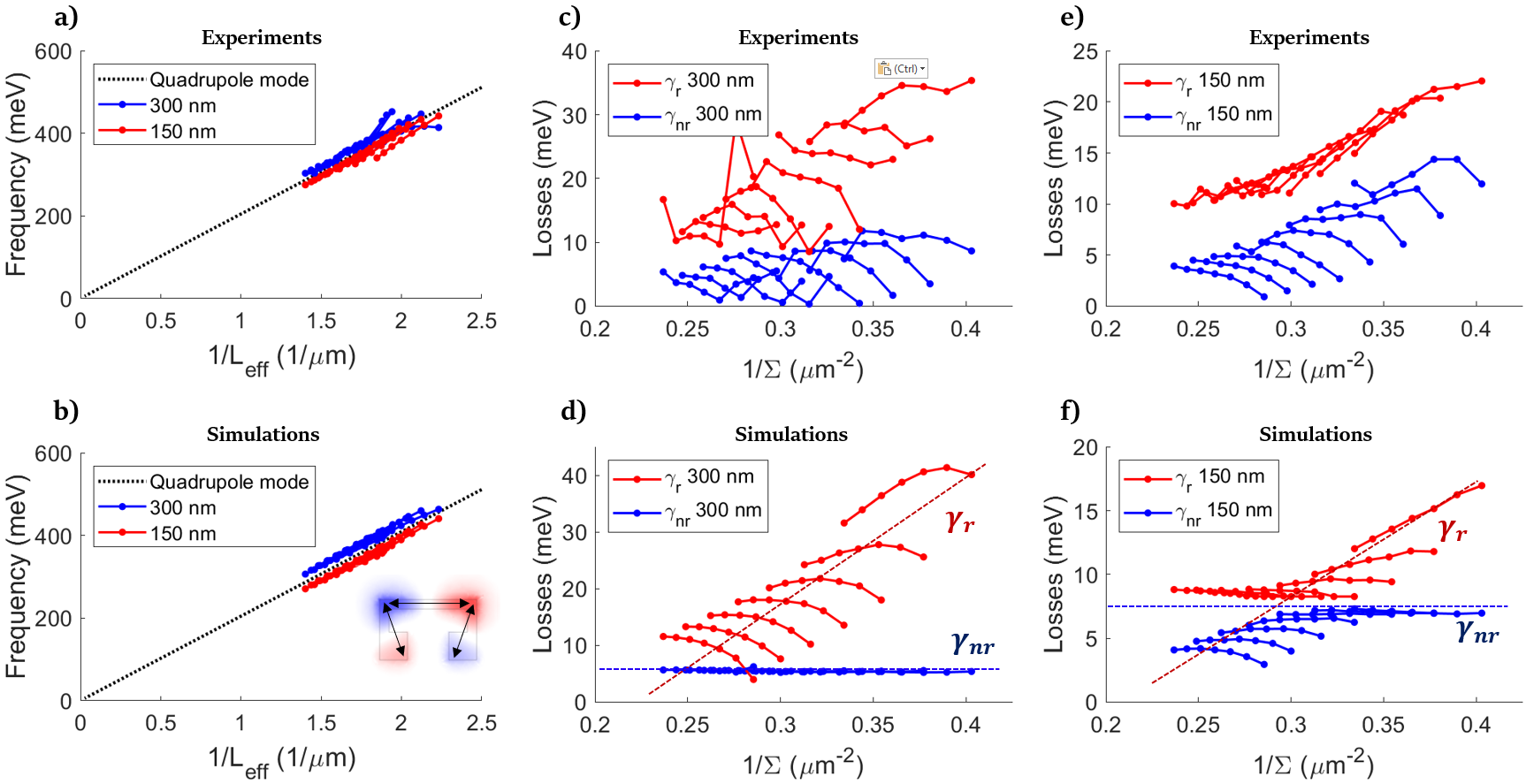}
    \caption{Resonant frequency of the quadrupole mode as a function of the effective length $L_{eff}$ a, b). Both experiments and simulations show a linear behavior. The dotted lines represent the resonant frequency of the quadrupolar mode of a patch with a $n_{eff}=3.5$. Radiative and non-radiative losses of the structure with a thickness $h=300$ nm (c, d) and $150$ nm (e, f) as a function of the array unit cell area $\Sigma$. The dashed lines serve as a guide for the eyes, highlighting the two type of losses.}
    \label{Fig8}
\end{figure}

\textbf{Quadrupole mode}. To complete the study, we analyzed the frequency response and losses of the quadrupolar mode (Figure \ref{Fig3}, d) and Figure \ref{Fig4}, e)). The resonant frequency of this mode follows that of a rectangular patch cavity \cite{Markad2015}: 
\begin{equation}
\omega_{Q_X} = \frac{2\pi c}{2n_{eff}} \left[ \frac{1}{L_x^2} + \frac{1}{L_y^2} \right]^{1/2}, 
\label{omegaQx}
\end{equation}
where $L_x$ and $L_y$ are the sizes along the $x$- and $y$-axis, respectively. The effective refractive index for this mode was evaluated with eq.\eqref{neff}: $n_{eff}\simeq3.55$ for the 150 nm structure and $3.1$ for the 300 nm one. These values are slightly lower than those found for the patch mode, which can be attributed to the reduced penetration of the electric field into the gold layer. This reduction occurs because the skin depth decreases at higher frequencies \cite{Rioux2014}.\\
As noted earlier, the field oscillations conform to the shape of the resonator. Specifically, the distance between the maxima in the capacitor and the edge of the wire is given by $\displaystyle L_y^* =\sqrt{(L_y + L_C)^2 + L_C^2}$ (inset in Fig.\ref{Fig8}, b). This expression allows us to define an effective length: $\displaystyle\frac{1}{L_{eff}} = \left( \frac{1}{L_x^2} + \frac{1}{L_y^{*2}} \right)^{1/2}$, which can then be substituted into eq.\eqref{omegaQx}. In figure \ref{Fig8} (a, b) we report the resonant frequency as a function of $1/L_{eff}$ for both thicknesses. The dotted line represents the plot of eq.\eqref{omegaQx} with an effective refractive index of $3.5$, included to highlight the linear behavior of the data. 

The losses are reported in Fig.\ref{Fig8} (c-f), showing a trend similar to that of the patch mode. Radiative losses scale linearly with the thickness of the samples, and we observe a crossing of the two loss channels corresponding to the critical coupling condition (Fig.\ref{Fig8}, e, f). Indeed, or all devices fabricated on the $150$ nm thick sample, the quadrupole mode exhibited nearly perfect absorption (see supporting information for the spectra). The analysis of the 300 nm sample (Fig.\ref{Fig8}, c, f) was more challenging due to the excitation of the lattice mode (a propagating surface plasmon) at high frequencies \cite{jouy2011coupling}. The two resonances overlapped and interact, increasing the observed width and shifting the measured resonant frequency.These effects rendered Lorentzian fitting impractical for certain spectra, and those points were excluded from Fig.\ref{Fig8}, c). Despite these complications, Fig.\ref{Fig8}, (d, f) clearly shows a consistent trend: the non-radiative losses remain approximately constant, while the radiative losses increase with the array unit area $\Sigma$. The dashed lines, representing radiative and non-radiative losses, are included as a guide for the eyes.

\begin{figure}[ht]
    \centering
    \includegraphics[width=1\textwidth]{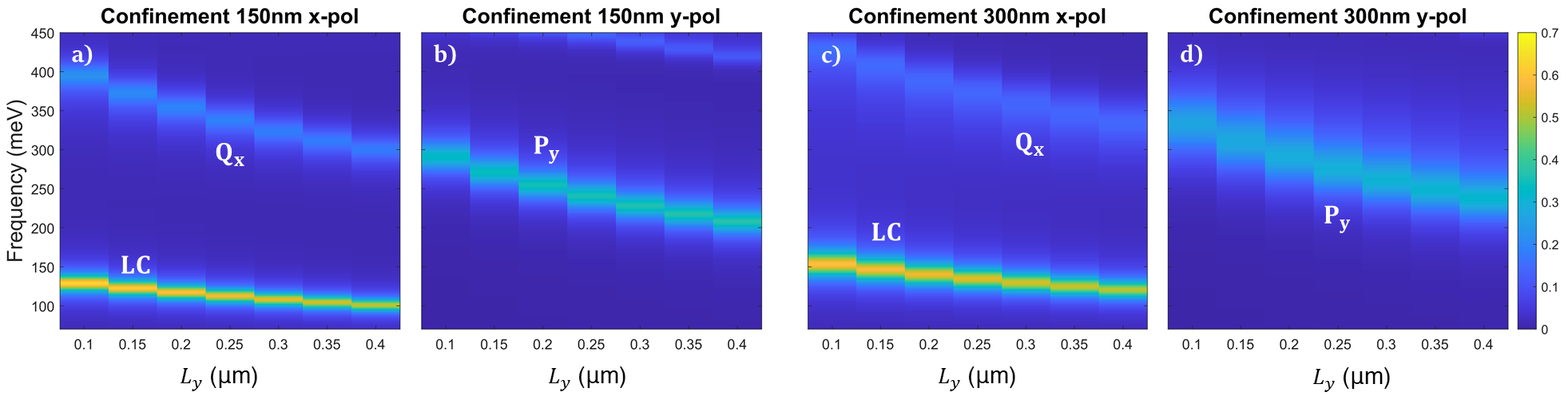}
    \caption{Confinement factor of the three modes obtained from simulations. The spectra are reported as a function of $L_y$ for a fixed  $L_x=1$ $\mu$m, for both polarization and thicknesses. The colormap is fixed for all the panels to the maximum value obtained ($\sim0.7$) to directly compare all the modes. The highest value is achieved by the LC mode for the thinnest structure (70\%). The patch and quadrupole mode reached 35\% and 20\% respectively.}
    \label{Fig9}
\end{figure}

\textbf{Confinement of the modes}. To conclude, we compare the confinement factor of the modes discussed above. The confinement factor measure the ability of the resonator to focus the electromagnetic energy in a small (sub-wavelengths) volume. Specifically, we are interested in the energy confined between the capacitors and the bottom metallic ground plane, within a volume $V_C=2L_C^2h$. To quantitatively evaluate this confinement factor, we calculated the electrical energy contained within this region, normalized by the total electrical energy in the entire simulation volume. Notably, in the context of utilizing the resonant mode for intersubband devices, we focused on the electrical energy from the $z$-component of the electric field:
\begin{equation}
c_{E_z} = \frac{ \int_{V_C} \varepsilon_r E_z^2 \, d^3x }{ \int_{V_{sim}} \varepsilon_r E^2 \, d^3x },
\end{equation}
where $V_{sim}$ is the whole simulation box volume. Despite some leakage through the side arms, the LC mode achieved values as high as 0.7, meaning that 70\% of the total energy within the simulation was concentrated in the $z$-component of the electric field below the capacitors. This volume is approximately $10^5$ times smaller than $\lambda_0^3$, highlighting the power of such resonators to significantly enhance the field intensity inside the material. In Figure \ref{Fig9}, we present the confinement factor as a function of frequency for the two thicknesses and polarizations. Again, we limited the plots to one row of the matrix: $L_x=1$ $\mu$m while $L_y$ varies. The color scale was fixed for all the panels (from 0 to 0.7), allowing for direct comparison of the modes. The highest value is achieved by the LC mode in the thinnest structure. This trend can be attributed to greater leakage in thicker structures, particularly through the side facets into the surrounding air.\\
The patch and quadrupolar mode reached respectively $35\%$ and $20\%$. These lower values can be readily understood by examining the field distribution shown in Figure \ref{Fig3}: both modes extend across the entire volume of the resonator and are not confined by the capacitive region, further emphasizing the physical distinction between the quasi-static LC mode and the propagating modes. It is important to stress again that here we plotted the confinement factor calculated for the volume comprised below the capacitors to highlight the behavior of the LC mode. By calculating the confinement factor in the whole resonator volume, we achieved values around 85\% for all modes, highlighting their strong potential for device applications.\\

\section{Conclusions}
We presented an in-depth analysis of the resonant modes of three-dimensional meta-atom resonators in the Mid-IR, focusing on their experimental characterization and theoretical modeling. The extremely short wavelength posed significant challenges due to the increased metal losses and complex fabrication procedures. To the knowledge of the authors, this work represents the first experimental characterization of these three-dimensional meta-atoms in the mid-IR range.\\
The analytical model we developed allows to quantitatively predict the resonant frequency of the modes as well as radiative and non-radiative losses. This model can be in principle generalized to any resonator shape, granting a powerful tool to achieve the desired electromagnetic response. In particular, the design we proposed demonstrated an almost independent tunability of the various mode frequencies and losses, making these structures highly versatile. In example, this capability offers the opportunity to explore non-linear effects \cite{Lee2014,yu2019third}. Indeed, it is possible to design a resonator where the patch and quadrupolar modes resonate at frequencies that are respectively twice and three times that of the main LC mode. Furthermore, owing to the high confinement factor and strong optical absorption, the field intensity within the material is significantly enhanced, thereby amplifying non-linear interactions.\\
These resonators are not only useful for studying non-linear effects but also show great promise for Mid-IR photodetectors. The LC mode, which confines light to extremely small volumes, offers a unique advantage by reducing the electrical area while maintaining high absorption. This is crucial to minimize dark currents and capacitive effects, allowing higher temperature operation and faster detection speeds \cite{Jeannin2020a}. 
The extreme confinement factor of these resonators also makes them ideal candidates to investigate the regime of strong light-matter coupling in the few-electron limit that could be reached for such small sizes \cite{todorov2014few}.\\
The combination of our three-dimensional meta-atom design and analytical model provides a versatile platform with a high degree of tunability, achieved by varying a few key geometrical parameters, that will be an invaluable tool to explore a wide range of phenomena.\\
\textbf{Acknowledgment}: The authors thank the Renatech network for the sample growth; we acknowledge funding from ERC-COG-863487 UNIQUE.

\section{Methods}
\textbf{Sample geometry}. As shown in Fig.\ref{Fig1}, the resonators are fabricated in a metal-dielectric-metal configuration. The bottom metal consists of a 150 nm gold (Au) layer, deposited on top of a 10 nm titanium (Ti) layer, which is introduced through wafer bonding. The Ti layer ensures good adhesion of the Au. The top metal layer, made of 10 nm Ti and 100 nm Au, is patterned into a $\Pi$ shape using e-beam lithography. The structure consists of a rectangular loop with dimensions $L_x$ and $L_y$ (see Fig.\ref{Fig1}(a)), terminated by two square pads of size $L_C$. The dielectric layer is composed of Gallium Arsenide (GaAs). After fabrication, an etch step using inductively coupled plasma (ICP) is performed to expose the bottom gold layer.
Our goal is to systematically investigate how the resonator shape influences the electromagnetic resonances supported by the structure. To achieve this, we fabricated multiple arrays with varying geometrical parameters for the top wire ($L_x$ and $L_y$), as illustrated in Fig.\ref{Fig1}(a), while keeping the capacitor size constant at $L_C=300$ nm. Each array forms a square lattice with a total area of $100 \times 100 , \mu$m$^2$, comprising resonators of identical shape ($L_x$ and $L_y$), spaced 1 $\mu$m apart in both directions. The full structure forms a matrix with $8 \times 7$ different geometrical combinations, where $L_x$ ranges from 0.7 to 1.4 $\mu$m, and $L_y$ ranges from 0.1 to 0.4 $\mu$m. Additionally, a separate column was fabricated with varying capacitor sizes. To assess the impact of different structural thicknesses on the resonance, identical fabrications were performed using GaAs layers of two different thicknesses, $h=150$ nm and $h=300$ nm.

\textbf{Reflectance spectra}. To probe the electromagnetic modes of the structures, we performed spectrally resolved reflectivity measurements using a Fourier-transform infrared spectrometer coupled with an infrared microscope, enabling us to examine each array individually. This setup is referred to as "micro-FTIR." The measurements were carried out with the electric field polarized along both the $x$ and $y$ directions, and each spectrum was normalized by the reflectivity of an unpatterned gold area on the bottom of the structure. Typical results for both thicknesses, $h = 150$ nm and $h = 300$ nm, are presented in Fig.\ref{Fig2}.
It is important to note that, due to the Cassegrain design of the objective in the micro-FTIR setup, the light impinges on the sample within a cone, encompassing a range of angles between 12\textdegree and 24\textdegree. The reflectivity we measured results from the convolution of these angles. In contrasts in the simulations, we assumed a fixed angle of 0\textdegree. This effect contributes, to the broader measured linewidth. Moreover, again due to the angle of incidence, we observed the lattice mode resonance (surface plasmon polariton) at lower frequencies (400-500 meV) than simulated  high frequencies (500-600 meV). See supporting information.

\textbf{Numerical simulations}. Simulations of the reflectance were carried out using a commercial finite element software (COMSOL), which computes the electromagnetic response of the periodic arrangement of meta-atoms as a function of frequency. A port is used to introduce light into the structure and calculate the reflectivity. The geometry is discretized into a mesh grid, where Maxwell's equations are solved. The mesh is refined sufficiently to resolve the smallest element of the structure, in this case, the 100 nm wire. An adaptive algorithm adjusts the grid size for larger features, optimizing the computational time. Throughout the simulation domain, the maximum step size was set to be smaller than $\lambda/5n$, where $\lambda$ is the wavelength and $n$ is the refractive index of the material.\\

\textbf{SNOM measurements.} We confirmed the field distribution shown in the simulations using an attocube-Neaspec Scanning Near Optical Microscope (SNOM). A tunable laser (MIRcat) was directed onto an uncoated silicon tip positioned above the resonators. The light impinged at a 60\textdegree incidence angle and was vertically polarized. The electric field interacted with the structure and was subsequently scattered toward a detector, as shown in Fig.\ref{Fig4}, a). The tip oscillated at 277 kHz with an amplitude of 80 nm. By analyzing the harmonics of the re-scattered light using a pseudo-heterodyne interferometer \cite{ocelic2006pseudoheterodyne,cvitkovic2007analytical}, we were able to reconstruct the amplitude and phase (second and third-order demodulation) of the field on top of the structure. To prevent plasmonic coupling between the meta-atoms and the tip, which would distort the field distribution, we avoided using metallic tips.
The resonator's size and orientation were selected so that the resonant absorption coincided with the laser emission and polarization. Specifically, the sample was rotated 90\textdegree to excite the patch mode (Fig.\ref{Fig4}, d). The raw phase data obtained from the SNOM measurements contained inherent discontinuities where the electric field changes sign, due to the periodic nature of the phase (ranging from $-\pi$ to $+\pi$). To address this, a phase unwrapping algorithm was applied to remove the $2\pi$ phase jumps and reconstruct the electric field distribution of Fig. \ref{Fig4}, c-e). The amplitude data displayed some deviations from the simulations, with the near-field distribution appearing strongly asymmetric within the resonator (see Supporting Information). This asymmetry is attributed to the tip altering the coupling with the far field, as discussed in \cite{thomas2022imaging}. To accurately reconstruct the field distribution, it would be necessary to perform multiple scans of the same resonator, rotating it incrementally within the plane (e.g., acquiring images at 0\textdegree, 45\textdegree, 90\textdegree, 135\textdegree, etc.). This approach would average out the asymmetry and provide a more accurate representation of the resonator's field distribution.

\newpage

\bibliographystyle{unsrt}
\bibliography{references}

\end{document}